\documentclass[sn-mathphys,Numbered]{sn-jnl}

\usepackage{graphicx}%
\usepackage{multirow}%
\usepackage{amsmath,amssymb,amsfonts}%
\usepackage{amsthm}%
\usepackage{mathrsfs}%
\usepackage[title]{appendix}%
\usepackage{xcolor}%
\usepackage{textcomp}%
\usepackage{manyfoot}%
\usepackage{multirow}%
\usepackage{booktabs}%
\usepackage{changepage}
\usepackage{geometry}
\usepackage{algorithm}%
\usepackage{algorithmicx}%
\usepackage{algpseudocode}%
\usepackage{listings}%

\theoremstyle{thmstyleone}%
\theoremstyle{thmstyletwo}%

\theoremstyle{thmstylethree}%
\newcommand{\chA}[1]{{{\color{black}{#1}}}}
\newcommand{\chE}[1]{{{\color{black}{#1}}}}

\begin{document}

\title[Article Title]{Diagnosising Helicobacter pylori using AutoEncoders and Limited Annotations through Anomalous Staining Patterns in IHC Whole Slide Images}


\author*[1,2]{\fnm{Pau} \sur{Cano}}\email{pcano@cvc.uab.cat}

\author[3,4]{\fnm{Eva} \sur{Musulen}}\email{eva@quironsalud.com}
\equalcont{These authors contributed equally to this work.}

\author*[1,2]{\fnm{Debora} \sur{Gil}}\email{debora@cvc.uab.cat}
\equalcont{These authors contributed equally to this work.}

\affil*[1]{\orgdiv{Comp. Sci. Dep}, \orgname{ Universitat Autònoma de Barcelona}, \orgaddress{\street{campus UAB}, \city{Cerdanyola del Vallès}, \postcode{08193}, \state{Catalunya}, \country{Spain}}}

\affil[2]{\orgname{Computer Vision Center}, \orgaddress{\street{campus UAB}, \city{Cerdanyola del Vallès}, \postcode{08193}, \state{Catalunya}, \country{Spain}}}

\affil[3]{\orgdiv{Pathology Department}, \orgname{Hospital Universitari General de Catalunya-Grupo QuironSalud}, \orgaddress{\city{Sant Cugat del Vallès}, \postcode{08195}, \state{Catalunya}, \country{Spain}}}

\affil[4]{\orgname{Institut de Recerca contra la Leucèmia Josep Carreras (IJC)}, \orgaddress{\city{Badalona}, \postcode{08916}, \state{Catalunya}, \country{Spain}}}


\abstract{\textbf{Purpose:} This work addresses the detection of Helicobacter pylori (\textit{H. pylori}) in histological images with immunohistochemical staining.  This analysis is a time demanding task, currently done by an expert pathologist that visually inspects the samples. \chA{Given the effort required to localise the pathogen in images, a limited number of annotations might be available in an initial setting. Our goal is to design an approach that, using a limited set of annotations, is capable of obtaining results good enough to be used as a support tool.} 
 
\textbf{Methods:}   We propose to use autoencoders to learn the latent patterns of healthy patches and formulate a specific measure of the reconstruction error of the image in HSV space. ROC analysis is used to set the optimal threshold of this measure and the percentage of positive patches in a sample that determines the presence of \textit{H. pylori}.
 
\textbf{Results:} Our method has been tested on an own database of 245 Whole Slide Images (WSI) having 117 cases without \textit{H. pylori} and different density of the bacteria in the remaining ones. The database has 1211 annotated patches, with only 163 positive patches. This dataset of positive annotations was used to train a baseline thresholding and an SVM using the features of a pre-trained RedNet-18 and ViT models. A 10-fold cross-validation shows that our method has better performance with 91\% accuracy, 86\%  sensitivity, 96\% specificity and 0.97 AUC in the diagnosis of \textit{ H. pylori }.

\textbf{Conclusion:} Unlike classification approaches, our shallow autoencoder with threshold adaptation for the detection of anomalous staining is able to achieve competitive results with a limited set of annotated data. This initial approach is good enough to be used as a guide for fast annotation of infected patches. }




\maketitle

\section{Introduction}\label{sec1}

The bacterium \textit{Helicobacter pylori} (\textit{H. pylori}) is the main cause of gastritis, an inflammation of the gastric mucosa that can lead to other serious diseases, such as gastric ulcer and even cancer. It is ubiquitous and more than 50\% of the world's population has been infected by the bacterium, with a prevalence that exceeds 80\% in adults over fifty\cite{yang}. Early detection of this bacterium is essential for the effective diagnosis, treatment of infection and prevention of secondary pathologies. 

The diagnosis of \textit{H. pylori} is usually made by visualization of gastric mucosa samples \chA{in Whole-Slide Images (WSI)}. Although \textit{H. pylori} is visible in conventional H\&E stains, \chA{the use of more specific} histochemical techniques, such as Giemsa or Warthin-Starry (WS), facilitates its visualization.  Among them, immunohistochemical staining of the bacterium is the technique that allows the most accurate diagnosis \cite{Batts}. This technique allows the visualization of the bacterium through the staining of specific proteins present in its membrane. In this manner, \textit{H. pylori} is stained with a color different from the one of other tissue, which avoids false detection of \textit{H. pylori} due to other gram-negative bacteria present in the sample. \chE{Immunohistochemical staining with diaminobenzidine in combination with the secondary antibody, gives the specific protein of \textit{H. pylori} a brown colored precipitate at the site of specific antibody binding}, while the rest of the tissue remains in a bluish hue of the hematoxylin counter stain. \chA{Even if} this facilitates the visual identification of \textit{H. pylori}, a pathologist still must carefully inspect the whole immunohistochemistry images in order to identify areas with \textit{H. pylori}. Since the bacteria are mostly located at the borders of the tissue samples, the pathologists must carefully inspect a zoom-up area for all points belonging to the border. Given the enormous size of the images (120000x16000 pixels) and the fact that several sections of the tissue sample appear in one digitized image, this manual inspection is a highly time consuming task that becomes harder the lower the concentration of \textit{H. pylori} is.

\begin{figure}[!h]
\centering
	\includegraphics[width=0.9\columnwidth]{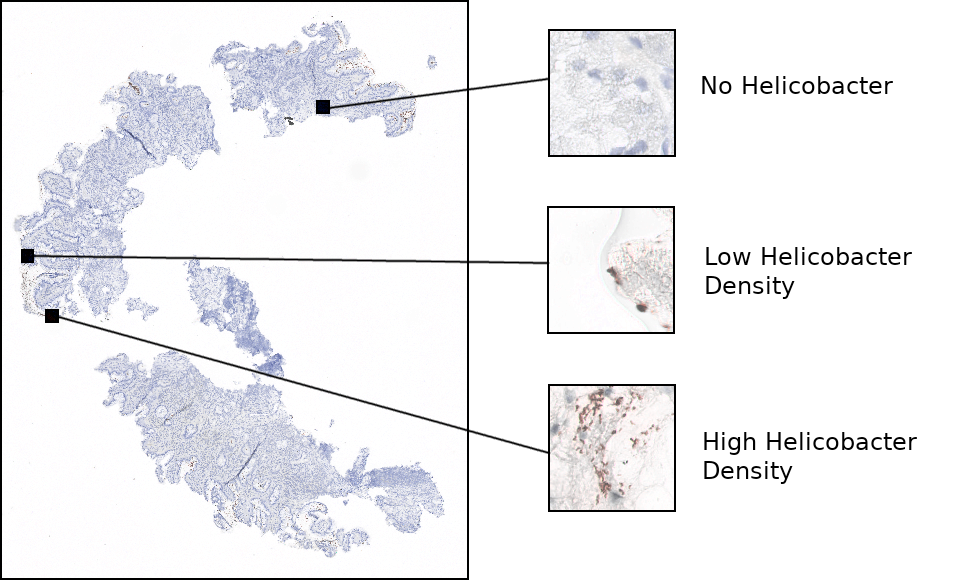}
	\caption{Histological sample with immunohistochemical staining. On the right: 3 windows of the same sample with different levels of \textit{H. pylori} density}
	\label{fig-histology-inmuno}
\end{figure}

Figure \ref{fig-histology-inmuno} shows a gastric mucosa section with positive immunohistochemical staining for \textit{H. pylori} and three close-up of tissue border regions with different density (negative, low and high) of the bacteria in the window images shown on the the right hand side. While the window with high presence of \textit{H. pylori}i is easily identified, the window with low density needs a more careful inspection in order to detect the brown colored spots of \textit{H. pylori} and avoid confusion with other artifacts that can be in the sample.


\chA{The recent development of WSI scanning has allowed the development of computer-assisted programs. In this work, we propose a method to automatically analyze \chA{WSI} of gastric tissue with immunohistochemical staining for the detection of
\textit{H. pylori}.}

\subsection{State-of-the-Art}
\label{SoA}

Although Deep Learning, DL, models have demonstrated good performance on several histopathologic tasks \cite{Salto}, there are not \chA{so} many works addressing the detection of \textit{H. pylori}. Existing works are based on \chA{different convolutional architectures} for the supervised classification of cropped patches extracted from tissue samples into \textit{H. pylori} positive or negative samples. 

In \cite{Klein} the authors trained a compact VGG-style architecture on, both, Giemsa and H\&E slides. The trained network was used to highlight regions of \textit{H. pylori} presence and tested as decision support system for the diagnosis of the bacteria. The network was able to classify Giemsa stained samples with a sensitivity of 1 with a low specificity of 0.66. In \cite{Liscia} the authors also used \chA{a VGG-like model for the classification of patches in silver staining samples. They achieved a sensitivity and specificity of, respectively, $0.89$ and $0.87$ 
with a $77\%$ of precision in the detection of patches with \textit{H. pylori}. Given that \textit{H. pylori} can concentrate on colonies at specific spots of the mucosa border and the size of WSI, a precision of $77\%$ might arise too many false positives to be used in a decision support system.} \chA{In \cite{Ibrahim}, the authors explored the potential of five 
pre-trained models (DenseNet-201, EfficientNet-b0, MobileNet-v2, ResNet-101, Xception) for binary classification of 204 H\&E-stained images cropped from WSI. The analysis of results using a 5-fold validation concluded that ResNet-101 was the best performer, folowed by DenseNet-201 with AUC of 0.9417 and 0.9383, respectively. There are not many works addressing the classification of WSI images for the diagnosis of \textit{H. pylori}.}

In \cite{Zhou}, the authors proposed an ensemble model of the output probabilities of 3 ResNet-18 and 3 DenseNet-121 models trained on patches cropped from H\&E-stained WSI. \chA{In this work,} patch-level probabilities were aggregated into WSI-level probabilities by averaging the top 10 patch-level probabilities of each section \chA{in order to issue a diagnosis of the sample}. This ensemble achieved a sensitivity of $0.87$, a specificity of $0.92$ and F1-score of 0.89 for the diagnosis of WSI. The model was also tested as support system to improve the performance of \chA{a pathologist during visual inspection of WSI}. \chA{The use of patch probabilities to guide visual inspection improved the accuracy when diagnosing \textit{H. pylori} positive samples, but introduced a higher uncertainty \chA{in case of} \textit{H. pylori} negative samples}.

\chA{A recent method reported in \cite{Lin} proposes a two-tier deep learning-based model for diagnosing \textit{H. pylori} in H\&E-staining that combines global analysis of WSI with local detection of the presence of the pathogen. The method bases on a weakly-supervised training to predict the diagnosis from analysis of WSI using the enhanced streamed CNN reported in \cite{Huang}. In spite of the impressive results achieved on a single test set split (AUC of 0.98), an analysis of the image areas (extracted using CAM \cite{CAM}) that the WSI model was using for predictions showed that they were (with average precision of 0.4044) areas of gastric mucosa tissue and not the surface epithelial cells or mucinous layers, which are the primary sites of \textit{H. pylori} colonization.  The authors concluded that the WSI model classified gastritis mainly based on inflammatory patterns and, thus, implemented an auxiliary model for bacilli localization because identifying HP bacilli is mandatory for diagnosing HP gastritis. This auxiliary model was trained on 824 patches (446 negatives) extracted from immunohistochemically stained images for the confirmation of \textit{H. pylori} and achieved an AUC of 0.95.}  



One of the main challenges for the use of supervised classification approaches for the identification of \textit{H. pylori} in \chA{WSI} images is the collection of enough annotated data, since this implies a time consuming visual inspection and identification of patches containing the bacteria. 

Self-supervised methods \cite{ContrastiveReview} allow learning representation spaces that capture abstract structures useful for performing downstream tasks. By their higher performance in task leveraging in problems with a small amount of annotated date, they are increasing their popularity in medical imaging problems. In particular, the latent space of generative models can be used to detect diseased tissue in histopathological images \cite{BreastCancer}. In particular, the authors used an autoencoder to learn a representation space preserving the manifold structure of input images to classify breast cancer. 

In this work, we pose the detection of \textit{H. pylori} as a detection of anomalies in the staining of tissue in immunohistochemically stained WSI by means of an autoencoder (AE).
By training the autoencoder with patches extracted from patients without \textit{H. pylori}, the latent space is a representation of non-infected blue tissue and windows with the presence of brown stained \textit{H. pylori} are poorly reconstructed. Since the autoencoder is not able to properly reconstruct the brown colored precipitate   associated to \textit{H. pylori} staining, we formulate a specific measure of this reconstruction error in HSV color space. This allows the detection of patches with \textit{H. pylori} as anomalies using very few annotated patches (of the order of hundreds!) in comparison to methods using the latent space as input for training a classifier \cite{BreastCancer}.

\section{Detection of \textit{H. pylori} using Autoencoders}\label{sec2}
\begin{figure*}[!h]
\centering
	\includegraphics[width=1\textwidth]{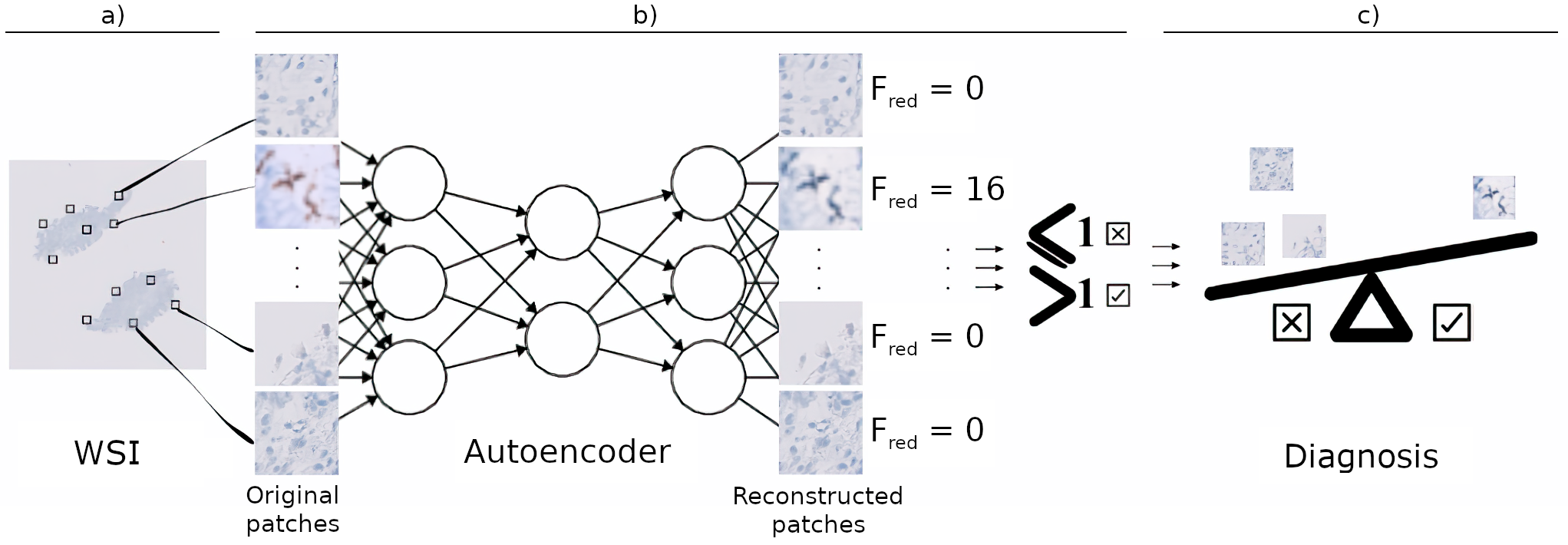}
	\caption{Schema of the main steps in the detection of H. pylori}
	\label{fig-autoencoder-architecture}
\end{figure*}

Our method has the following steps (sketched in  fig.\ref{fig-autoencoder-architecture}): detection of areas of interest in the image (fig.\ref{fig-autoencoder-architecture} a)), detection of anomalous stained elements in each region of interest (fig.\ref{fig-autoencoder-architecture} b)), and aggregation of each region of interest in the image for the diagnosis of the sample (fig.\ref{fig-autoencoder-architecture} c)). 
\chA{Given that \textit{H. pylori} is located along the
border, we use a morphological gradient of a tissue mask given by thresholding to detect the borders
of the tissue sample. Patches are defined by sliding windows of size 256x256 pixels cropped along pixels belonging to such borders. A set of windows extracted from healthy cases are the input to an AE for the reconstruction of the normal staining. The classification of patches into positive (\textit{H. pylori}) or negative is obtained using a metric based on the loss of brown-like pixels in the images reconstructed by the AE. The optimnal threshold for the metric is learned using a small set of annotated patches.} Finally, the percentage of positive windows defines a probability for the final classification of the sample.

\begin{figure}[!h]
\centering
\begin{tabular}{c|c}
  \includegraphics[width=0.4\columnwidth]{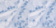}   & \includegraphics[width=0.4\columnwidth]{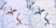} \\
     (a) & (b)
\end{tabular}
	\caption{Reconstructions of a healthy, (a), and infected, (b), patches. For each case, left images are the original inputs and right images, the reconstructions.}
	\label{fig-reconstruction-healthy}
\end{figure}

The AE is trained with patches from patients without \textit{H. pylori} in order to learn a representation space of normality (non-infected tissue) and detect \textit{H. pylori} as an anomalous staining. \chA{We propose to use a shallow architecture without bottleneck, since there are studies \cite{yong2022autoencoders} that show that non-bottlenecked architectures can be beneficial for anomaly detection}. In particular, the proposed AE has 3 convolutional blocks with one convolutional layer, batch normalization and leakyrelu activation. The size of the convolutional kernel is 3 and the number of neurons and stride of each layer are, respectively, [32,64,64] and [1,2,2]. Figure \ref{fig-reconstruction-healthy} shows the difference in the reconstructions of a non-infected patch, fig.\ref{fig-reconstruction-healthy}(a), and a patch with \textit{H. pylori}, fig.\ref{fig-reconstruction-healthy}(b). The reconstruction of the healthy patch looks like the original input image,  while the autoencoder has modified the coloration of the tissue in the reconstruction of the patch with \textit{H. pylori}.  In particular, the reconstruction has a color conversion to the blue hue and has lost the \chE{brown colored precipitate at the site of specific antibody binding} areas associated with the presence of \textit{H. pylori}. We use this difference in reconstructions to detect the presence of \textit{H. pylori} as follows. 

The presence of \textit{H. pylori} in a patch is computed using the fraction of brown colored pixels, $F_{brown}$, lost between original and reconstructed images. The brown colored pixels are computed applying a filter in HSV color space. In this color space, pixels with presence of \textit{H. pylori} have a hue in the range $[-20,20]$ and, thus, the area of red-like pixels is given by the number of pixels with hue in $[-20,20]$. If $H_{Ori}(i,j)$, $H_{Rec}(i,j)$ denote, respectively, the hue channel of the original and reconstructed patches, then $F_{brown}$ is formulated as:
\begin{equation}
    F_{brown}=\frac{\#\{(i,j) \mbox{ with } -20<H_{Ori}(i,j)<20\}}{\#\{(i,j) \mbox{ with } -20<H_{Rec}(i,j)<20\}}
\end{equation}\label{eq:Fred}

If $F_{brown}>1$, it indicates a loss of red-like pixels. \chA{The optimal value for $F_{brown}$ can be computed from a small number of annotated patches using ROC adaptive thresholding.} The percentage of patches in a tissue sample with $F_{brown}$ above this threshold defines the probability of presence of \textit{H. pylori}. The optimal threshold for this probability is obtained from the ROC curve as the probability of the closest point to $(0,1)$. Samples with a percentage of positive patches above this threshold are diagnosed as \textit{H. pylori} positive.

\section{Experiments}

\chA{The goal of these experiments is to show the advantages of our shallow AE for the detection of \textit{H. pylori} in WSI in case of a limited set of annotated patches. In order to do so, we have compared it to several approaches of different complexity: two pre-trained convolutional architectures and a simple thresholding on the amount of red pixels of the original patches. The pre-trained models are ResNet-18  (used in \cite{Zhou}) and the UNI Visual Transformer (ViT) model of \cite{UNIViT}. The ResNet-18 was pre-trained using natural images, while UNI was trained on 100,000 diagnostic H\&E-stained WSIs with different pathology across 20 major tissue types. For both architectures, we did a transfer learning and used their latent spaces as input to a SVM for the classification of patches. The baseline thresholding on red pixels was computed by the filter in HSV colour space used in (\ref{eq:Fred}) for $F_{brown}$. For all methods we used an adaptive thresholding on the percentage of positive patches computed using ROC analysis for the diagnosis of \textit{H. pylori}.}

\chA{Methods were tested on an own-collected} database of biopsies of antral or body gastric mucosa from the Department of Pathology of the Hospital Universitari General de Catalunya-Grupo Quironsalud. Formalin-fixed, paraffin-embedded tissue sections were analyzed using standard IHC techniques: immunostaining was performed automatically using a Ventana BenchMark ULTRA machine (\textit{Roche, Basel, Switzerland}) using the monoclonal primary antibody anti-Hp (\textit{clone SP48, Ventana Medical Systems, Inc., 1910 E. Innovation Park Drive, Tucson, Arizona 85755 USA}). An external positive control was included on each slide. All stained slides were scanned with an Ultra-Fast 180 slide scanner provided by Philips (\textit{Philips IntelliSite Pathology Solution}) to obtain WSI. 

The database has 245 WSI scored by a pathologist according to \textit{H. pylori} density as NEGATIVE (Healthy), LOW DENSITY and HIGH DENSITY. Of the 245 patients included in the study, 117 (47.8\% of the total) are classified as NEGATIVE, while 128 are classified as POSITIVE (LOW and HIGH DENSITY) for the presence of \textit{H. pylori}. \chA{Each WSI contains 2 sections of several gastric mucosa samples and one section extracted from an external positive control to check the validity of the immunohistochemical staining}.  We used the first diagnostic slide of the healthy cases to train models  and the second one to test the performance of the systems in the diagnosis of \textit{H. pylori}.  \chA{From a subset of 123 cases (77 positive ones), an expert pathologist annotated 1211 patches (161 being positive ones) extracted from the first slide.}

\chA{The set of annotated patches were used to train methods for the detection of the bacteria in patches. In particular, they were used to set the optimal threshold on $F_{brown}$ for our AE, the optimal threshold on red pixels and the training of SVM for the DL models. Regarding, AE training, for each healthy patient, 50 windows where randomly cropped from tissue borders of the first sample slide, which gives a total number of 5850 windows for training the AE model.}

\chE{The patches cropped from the digitalized WSI can be downloaded from the github repository https://github.com/IAM-CVC/HelicoBacterDetection, which also contains the code for training and testing our models.}

The performance metrics we have considered are the precision, recall and F-1 score for each diagnostic class (positive \textit{H. pylori} or negative \textit{H. pylori}). In order to allow for statistical assessment of the performance, the test set was split in 10 folds stratified by patient. For each fold, the optimal cutting point of the ROC curve was calculated from the training fold and tested in the independent set of patients.

\chA{Table \ref{tab:resultats-autoencoder} reports statistical summaries (average $\pm$ standard deviation) for the quality metrics obtained for each model. 
The proposed system has the highest specificity and F1-scores with average values of, respectively, 0.96 and 0.91. Comparing to a baseline thresholding on red pixels, the proposed AE approach significantly performs better for the detection of the negative class (No \textit{H. pylori}) with higher average recall (0.96 in comparison to 0.89) and more reproducible stable scores across folds (detected by the increase in standard deviations). The main reason for this drop in performance for healthy patients is that non infected tissue has also some red component in its staining. This baseline red component is learned by the AE model and, thus, only the reddish staining associated to H.pylori is poorly generated in the reconstructed patches. Comparing to more complex DL approaches, although the UNI ViT model clearly outperforms ResNet-18, none of them performs better than our AE architecture. In particular, our method has a higher specificity with lower standard deviation. We consider that a main reason for this drop in reproducibility can be the low number of positive annotated patches which is not large enough for a classification using a feature space of large dimensionality (1024 for the UNI ViT model and 512 for ResNet18). This low number of samples could also explain why the DL approaches perform worse than simple approaches (AE and red channel thresholdind). Finally, the worst performer is ResNet-18 with an average sensitivity of 0.68, in comparison to sensitivities above 0.8 for the remaining methods. This could be attributed to the domain of the images used for training ResNet-18, which, being natural images, is quite different from the histolgical samples used to train UNI ViT}


\begin{table}[h]
\caption{Statistical Summary of the 10-fold Validation}
\label{tab:resultats-autoencoder}
\centering
\begin{tabular}{lcccc}
   { }&{ }& \bf{No \textit{H. pylori}} &  \bf{\textit{H. pylori}} & \bf{Average} \\
  \cmidrule(lr){1-5}
  \multirow{3}{*}{\bf{Proposed AE}} & Precision & 0.86 $\pm$ 0.1 & 0.96 $\pm$ 0.07 & 0.91\\
  \cmidrule(lr){2-5}
   & Recall & 0.96 $\pm$ 0.09 & 0.86 $\pm$ 0.13 & 0.91 \\
  \cmidrule(lr){2-5}
   & F-1 score & 0.91 $\pm$ 0.06& 0.90 $\pm$ 0.07 & 0.91 \\
  \cmidrule(lr){1-5}
  \multirow{3}{*}{Baseline Threshold} & Precision & 0.85 $\pm$ 0.16 & 0.91 $\pm$ 0.09 & 0.88\\
  \cmidrule(lr){2-5}
   & Recall & 0.89 $\pm$ 0.13 & 0.84 $\pm$ 0.07 & 0.86 \\
  \cmidrule(lr){2-5}
   & F-1 score & 0.84 $\pm$ 0.12 & 0.91 $\pm$ 0.06 & 0.87 \\
  \cmidrule(lr){1-5}
  \multirow{3}{*}{ResNet-18} & Precision & 0.79 $\pm$ 0.14 & 0.73 $\pm$ 0.21 & 0.76\\
  \cmidrule(lr){2-5}
   & Recall & 0.80 $\pm$ 0.16 & 0.68 $\pm$ 0.21 & 0.74 \\
  \cmidrule(lr){2-5}
   & F-1 score & 0.79 $\pm$ 0.13& 0.69 $\pm$ 0.17 & 0.74 \\
  \cmidrule(lr){1-5}
  \multirow{3}{*}{UNI ViT} & Precision & 0.90 $\pm$ 0.12 & 0.80 $\pm$ 0.12 & 0.85\\
  \cmidrule(lr){2-5}
   & Recall & 0.82 $\pm$ 0.13 & 0.87 $\pm$ 0.16 & 0.84 \\
  \cmidrule(lr){2-5}
   & F-1 score & 0.86 $\pm$ 0.11& 0.82 $\pm$ 0.11 & 0.84 \\
  \cmidrule(lr){1-5}
\end{tabular}%
\end{table}

\chA{Table \ref{tab:confusion-matrix-autoencoder} shows the confusion matrix of the diagnosis of the 245 patients. Out of them, only 23 cases have been incorrectly classified by the proposed AE. This endows our method with higher accuracy in comparison to the alternative approaches that have 30, 24 and 51 miss-classifications for, respectively, baseline thresholding, UNI ViT and ResNet-18. The largest differences are for detection of healthy cases and, as before, ResNet-18 is the worst performer.}

\hspace*{-1.5cm}
\begin{table}[h]
\caption{Confusion Matrix of the 10-fold Validation }
\label{tab:confusion-matrix-autoencoder}

\centering

\begin{tabular}{lccclccc}
  { }&{ }& Predicted & Predicted &{ }&{ }& Predicted & Predicted\\
   { }&\multirow{1}{*}{\bf{Proposed AE}}&  \textit{H. pylori} &  No \textit{H. pylori} &{ }&\multirow{1}{*}{ResNet18}&  \textit{H. pylori} &  No \textit{H. pylori}\\
  \cmidrule(lr){2-4}\cmidrule(lr){6-8}
  { } & \textit{H. pylori} & 110 (TP) & 18 (FP) &{ }& \textit{H. pylori} & 87 & 41 \\
  \cmidrule(lr){2-4}\cmidrule(lr){6-8}
   & No \textit{H. pylori} & 5 (FN) & 112 (TN) &  & No \textit{H. pylori} & 10 & 107 \\
   \cmidrule(lr){2-4}\cmidrule(lr){6-8}
   \\
   { }&{ }& Predicted & Predicted &{ }&{ }& Predicted & Predicted\\
    { }& \multirow{1}{*}{Baseline Threshold}&  \textit{H. pylori} &  No \textit{H. pylori} & { }&\multirow{1}{*}{UNI ViT} &  \textit{H. pylori} &  No \textit{H. pylori}\\
  \cmidrule(lr){2-4}\cmidrule(lr){6-8}
  { } & \textit{H. pylori} & 118 & 10 & { }
  & \textit{H. pylori} &  115 & 13 \\
  \cmidrule(lr){2-4}\cmidrule(lr){6-8}
   & No \textit{H. pylori} & 20 & 96 &  & No \textit{H. pylori} &  21 & 96 \\
  \cmidrule(lr){2-4}\cmidrule(lr){6-8}
\end{tabular}%
\end{table}

\begin{figure}[!h]
\centering
\begin{tabular}{c}
	\includegraphics[width=0.95\columnwidth]{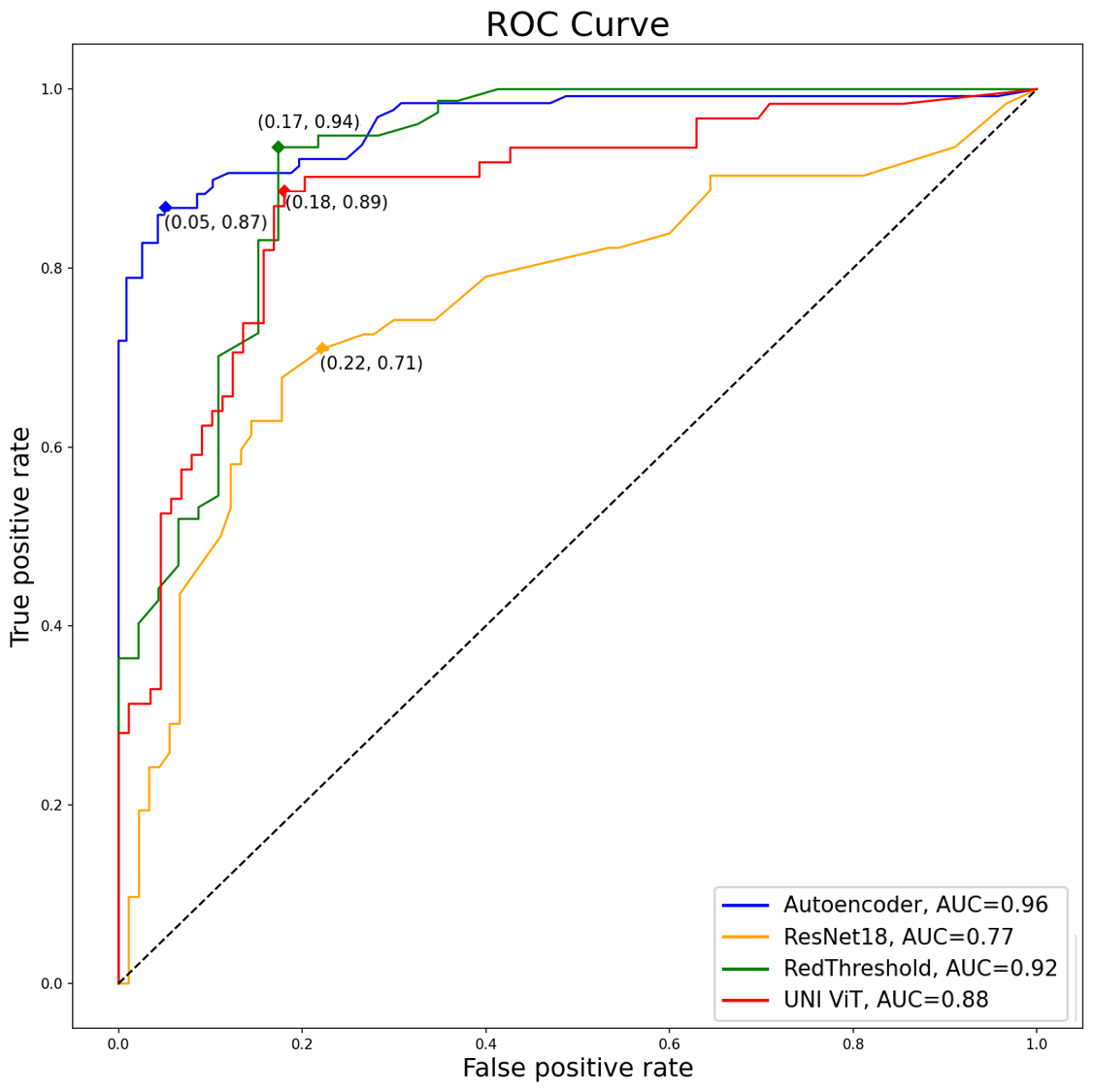}

\end{tabular}
	\caption{ROC curve of the 4 methods averaged for the 10 folds}
	\label{fig-roc-autoencoder}
\end{figure}

\chA{Finally, fig.\ref{fig-roc-autoencoder} shows the average ROC curve for the 4 methods with the point defining the optimal threshold for diagnosis. The proposed AE has the best ROC curve with an average AUC of 0.961. The second best is the baseline thresholding with an AUC of 0.92, followed by UNI ViT (AUC of 0.88) and ResNet-18 being the worst approach with AUC of 0.77. It is noticeable the stability of AE cutting point across folds, with a variability of 0.1 in the ranges (6.18\% $\pm$ 0.10\%) of the thresholding value of the probability of  \textit{H. pylori}.} 


\section{Conclusion}\label{sec13}

We have presented a DL system for the diagnosis of \textit{H. pylori} on immunohistochemically stained gastric mucosa based on autoencoders trained to obtain a normality pattern from non-infected samples.  \chE{ We have also released the code and the patches (including annotated ones and the diagnosis for each patient) used in our experiments in github \footnote{https://github.com/IAM-CVC/HelicoBacterDetection}. As far as we know, this is the first publicly available dataset with patch annotations for the detection of \textit{H. pylori} on immunohistochemically stained gastric mucosa.}  

Autoencoders are able to detect \textit{H. pylori} as an anomaly in staining in a self-learning approach that does not require annotation of a large number of image patches. In particular, our initial dataset of own collected cases had only 163 annotated positive patches. This motivated the use of a shallow AE as an anomaly detector to obtain a first solution able to achieve competitive good results in such a limited set of annotated patches.  

The results of a comparison to a baseline thresholding of red pixels and task transfer using two different pre-trained architectures (ResNet-18 and UNI ViT) show that our AE has a much higher specificity (0.96) with similar sensitivity. A high specificity is a clinical requirement to avoid unnecessary treatments. The main reason for the competitive results of our AE compared to more complex approaches is that while the amount of positive annotations is large enough to compute an adaptive thresholding for the classification of patches, it is too short for training a classifier on the higher dimensional feature space given by the latent space of a CNN.

Comparing to other classification approaches working with other kinds of staining, our AE has better metrics: higher specificity compared to the 0.92 of \cite{Zhou} and higher AUC compared to the 0.95 of the auxiliary model of \cite{Lin}. Also comparing to self-supervised approaches in other histhopathological diagnostic areas leveraging the latent space, our measure of anomalous staining reconstruction is able to achieve similar results with a data set with $97\%$ less samples (245 in contrast to 7909 in \cite{BreastCancer}).

A main advantage of our approach is that it can produce competitive good metrics using a tiny number of annotated patches in a shallow architecture, thus, reducing the manual cost and time of the implementation. \chE{In particular, our method trains a shallow AE that generates healthy tissue staining from healthy samples and an adaptive threshold computed using ROC analysis that needs a set of pathological annotated patches and patients.}

\chE{Regarding AE, the minimum number of cases depends on the variability of the staining and morphology of tissue samples and it can only be empirically estimated through the analysis of the reconstructed images. This can be visually assessed and quantified using reconstruction metrics measuring the difference between input and output images. In particular, a number of samples could be considered optimal when adding more cases there are not significant differences on the reconstruction metrics in a k-fold cross-validation.}

\chE{Concerning sample size for the ROC analysis used to compute the adaptive thresholding, it can be computed using the formula for the estimation of the required sample size in the comparison of the area under a ROC curve with a null hypothesis value \cite{hanley1982meaning,obuchowski1997sample}.  The formula estimates the minimum number of samples required to achieve a given power and confidence level when testing if the expected empirical AUC is significantly above a minimum value ($>$0.5) in accuracy for \textit{H. pylori} detection. The power corresponds to  the TN rate of the test and it is related to the minimum difference between the two AUCs that the test considers to be significant. In our case, the number of samples in the cross-validation set for diagnosis allows to estimate differences of 0.07 in AUCs for comparisons to a baseline AUC=0.87 at 0.95 confidence with power of 0.8.}

\begin{figure}[!h]
\centering
	\includegraphics[width=0.85\columnwidth]{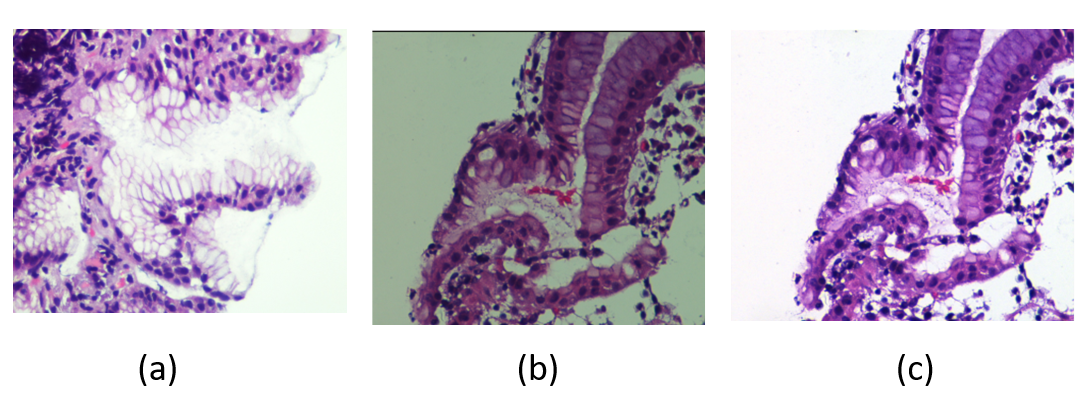}
	\caption{Example of Colour Transfer: target colour image, (a), source image, (b), transformed image, (c).}
	\label{fig-colortransfer}
\end{figure}

\chE{We are aware that without testing the method on other datasets might raise some concerns about the introduction of an over-fitting and the reproducibility of the approach. In this context, the techniques used for IHC are highly reproducible because they are automated with equipment and reagents provided by multinational companies such as Roche, Leica or Agilent. Given that our method bases on the detection of anomalies in staining, this standardization of the acquisition protocol highly guarantees the reproducibility of results on other datasets acquired using the same IHC staining (diaminobenzidine). In case of a bias in the baseline staining of images, colour transfer techniques like the ones reported in \cite{plenopticam} in conjunction with classical histogram matching can be used to match staining colours to the ones of our database.  Figure \ref{fig-colortransfer} illustrates the colour transfer for two images with H\&E staining. After applying the Multi-Variate Gaussian Distribution transfer reported in \cite{plenopticam} followed by a correction of the brightness in HSV space the colours of the image with biased baseline staining of fig.\ref{fig-colortransfer}, (b), match the target staining of the image shown in fig.\ref{fig-colortransfer}, (a).} 

\chE{We are currently working on several improvements. We are enlarging the set of annotated patches in a semi-supervised manner using the $F_{brown}$ score as a probability of having \textit{ H. pylori } to rank the patches of positive cases and show them to pathologists for the selection of those having the bacteria using a web application.} Concerning, the methodology, our current work focuses on the benefits of contrastive losses (like the normalized temperature-scaled cross entropy loss \cite{pmlr-v119-chen20j}) for improving the detection in immunohistochemical images and leveraging the task to H\&E stains images using domain adaptation approaches \cite{Wilm2022MindTG}. \chA{ We also consider that weakly-supervised analysis of WSI like the model presented in \cite{Lin} (with an impressive 0.98 of AUC for the WSI model detecting inflammation) could detect morphological changes associated to the presence of \textit{H. pylori} that could help to push our metrics. Thus, we also plan to investigate the benefits of tiered models integrating local and global information.} \chE{Finally, we are awaiting the recommendations of the Legal Cabinet of the Hospital on how to make the raw images in .tiff format public so as not to violate any data privacy law. We hope to have it before the end of the year. A link to the whole dataset will be added to the github release once approved by the Legal Cabinet of the Hospital. As far as we know, this will be the first publicly available dataset with, both, raw images and patch annotations for the detection of \textit{H. pylori} on immunohistochemically stained gastric mucosa.}

\backmatter

\bmhead{Acknowledgments}

This project is supported by the Ministerio de Economía, Industria y Competitividad, Gobierno de España grant number  PID 2021-126776OB-C21, Agència de Gestió d'Ajuts Universitaris i de Recerca grant numbers 2021SGR01623 and CERCA Programme / Generalitat de Catalunya.

\section*{Declarations}
\section*{Funding}
This project is supported by the Ministerio de Economía, Industria y Competitividad, Gobierno de España grant number  PID 2021-126776OB-C21, Agència de Gestió d'Ajuts Universitaris i de Recerca grant numbers 2021SGR01623 and CERCA Programme / Generalitat de Catalunya.

\section*{Conflict of interest}
The authors have no conflicts of interest to declare that are relevant to the content of this article.

\section*{Ethics approval}
This article does not contain any studies with human participants or animals performed by any of the authors.

\section*{Consent to participate}
This articles does not contain patient data


\bibliography{references}

\end{document}